\newif\ifabstract
\abstracttrue
 \abstractfalse 
\newif\iffull
\ifabstract \fullfalse \else \fulltrue \fi

\documentclass[11pt]{article}
\usepackage{amsfonts}
\usepackage{amssymb}
\usepackage{amstext}
\usepackage{amsmath}
\usepackage{xspace}
\usepackage{theorem}
\usepackage{graphicx}
\usepackage{url}
\usepackage{graphics}
\usepackage{colordvi}
\usepackage{colordvi}
\usepackage{subfigure}

\usepackage{natbib}
\usepackage{wasysym}

\usepackage{moreverb} 
\usepackage{textcomp} \usepackage{epsfig}
\usepackage{color} \usepackage{upgreek}
\usepackage{bbm}

\usepackage{subeqnarray}

\newcommand{\overbar}[1]{\mkern 1.5mu\overline{\mkern-1.5mu#1\mkern-1.5mu}\mkern 1.5mu}

\advance \oddsidemargin by -0.8in
\newcommand{\myparskip}{3pt}
\parskip \myparskip

\newsavebox{\astrutbox}
\sbox{\astrutbox}{\rule[-5pt]{0pt}{20pt}}

\begin{document}

\title{Heat and momentum transport in turbulent horizontal convection at low Prandtl numbers}
\author{Pierre-Yves Passaggia\thanks{Department of Marine Sciences, University of North Carolina, Chapel Hill, NC 27599, USA. Email: {\tt passaggia@unc.edu}.}
\and Alberto Scotti\thanks{Department of Marine Sciences, University of North Carolina, Chapel Hill, NC 27599, USA. Email: {\tt ascotti@unc.edu}.} 
\and Brian L. White\thanks{Department of Marine Sciences, University of North Carolina, Chapel Hill, NC 27599, USA. Email: {\tt bwhite@unc.edu}.}}

\maketitle

\thispagestyle{empty}

\begin{abstract}
The transition to a new turbulent regime in horizontal convection in the case of low Prandtl numbers is analyzed using the Shishkina, Grossmann \& Lohse (SGL) theory. The flow driven by the horizontal gradient along a horizontal surface, perpendicular to the acceleration of gravity is shown to transition to turbulence in the plume and the core. This transition to turbulence sets a sequence of heat transfer and momentum transport scalings which are 
found to follow the SGL prediction for the scaling factors and the prediction of Hughes, Griffith \& Mullarney (HGM) for larger forcing amplitudes. These results embed the HGM model in the SGL theory, agreed and extends the known regime diagram of horizontal convection, and provide the first evidence of both regimes at low and intermediate Prandtl numbers and sheds new insights on the role of HC in the earth's inner core dynamics. 
\end{abstract}

Turbulent Horizontal Convection (HC) in highly thermally conductive liquids plays a major role in geophysical flows. For example, it drives the Earth's outer core dynamics via either heat transport  from the inner to the outer core \cite{bloxham:91} or Joule effect due to Earth's magnetic field \cite{takehiro:11}. Although a lot of attention has been devoted to Rayleigh-B\'enard Convection (RBC) for the outer core's dynamics, it is only very recently that HC has attracted the attention of planetary scientists \cite{alboussiere:12}.
Lateral motions in the outer core are responsible for driving large scale horizontal flows in the mantle and have long been thought to be at the origin of striping and faulting of tectonic plates and strike-slip earthquakes \cite{Knopoff63}. At the edge of Earth's inner core, horizontal regions of thermally stable (crystallizing) and unstable (melting) stratified layers explain the East-West asymmetry of the inner core\cite{alboussiere:12}. However, only very little is known about the properties of the turbulent horizontal flows generated in these regions.
In this letter, we report Direct Numerical Simulation (DNS) results on how
 the Reynolds number ($\rm{Re}$), the turbulent kinetic energy dissipation ($\epsilon_u$) and the Nusselt number ($\rm{Nu}$) depend on the Rayleigh number ($\rm{Ra}$) and the Prandlt number ($\rm{Pr}$) in
turbulent HC at low $\rm{Pr}$ for different Prandtl numbers characteristic of convection in gases where $0.1<\rm{Pr}<1$ \cite{Roche02,Taylor13},
and liquid metals where $\rm{Pr}=\mathcal{O}(10^{-2})$ (see ref.\cite{takehiro:11}). The results are in agreement with the scaling power laws recently derived by Shishkina {\it et al.} \cite{ShishkinaGL16} based on the original work of Grossmann \& Lohse 
\cite{GL00} (GL) and numerical simulations of Takehiro \cite{takehiro:11}. It also provides a connection between the GL theory and the plume driven dynamics derived by Hughes {\it et al.} \cite{Hughes07}. 
Our simulations cover the laminar Rossby regime $I_l$ (see ref.\cite{Rossby65}), the high-$\rm{Pr}$ laminar regime $I^*_l$ recently reported by Shishkina  \& Wagner \cite{ShishkinaW16} and a new low-$\rm{Pr}$ turbulent regimes named $II_l$ (see ref.\cite{shishkina2017scaling,GL00} for theoretical predictions of HC and RBC), which is the first turbulent limiting regime reported in HC. We also observe the plume dominated flow regime of Hughes {\it et al.} (see ref.\cite{Hughes07}), that we name $II_u$ according to the SGL theory. These results also agree and extend the regime diagram of horizontal convection proposed in Hughes \& Griffith (see ref.\cite{hughes2008horizontal}).\\\\

Similarly to Shishkina \& Wagner \cite{ShishkinaW16}, we exploit the idea that in turbulent thermal convection, the time- and volume-averaged thermal and viscous dissipation rates are determined to leading order by their bulk or Boundary Layer (BL) contributions. For the ease of comparison, we follow the same way of presenting as Shishkina \& Wagner \cite{ShishkinaW16}.
We consider here the problem of convection in the Boussinesq limit, where the density difference $\Delta\rho=\rho_{max}-\rho_{min}$ across the horizontal surface is a small deviation from the reference density $\rho_{min}$. In this limit, the equations of fluid motion are
\begin{subeqnarray}
\frac{D \mathbf{u}}{D t} &=& -\nabla p + b\mathbf{e_z}  + \left(\frac{\rm{Pr}}{\rm{Ra}}\right)^{1/2}\nabla^2 \mathbf{u} \\
\nabla\cdot\mathbf{u}&=&0 \\
\frac{D b}{D t} &=& \left(\rm{Pr}\,\rm{Ra}\right)^{-1/2}\nabla^2 b,
\label{NS}
\end{subeqnarray}
where $D/Dt$ denotes the material derivative, $\mathbf{u}=(u,v,w)^T$ is the velocity vector, $b=-g(\rho-\rho_{min})/\rho_{min}$ is the buoyancy, $g$ is the acceleration of gravity along the vertical unit vector $\mathbf{e}_z$ and $p$ is the hydrodynamic pressure.
\begin{figure*}[t!]
\hspace{-0mm}
\hspace{-2mm}\scalebox{0.185}{\Huge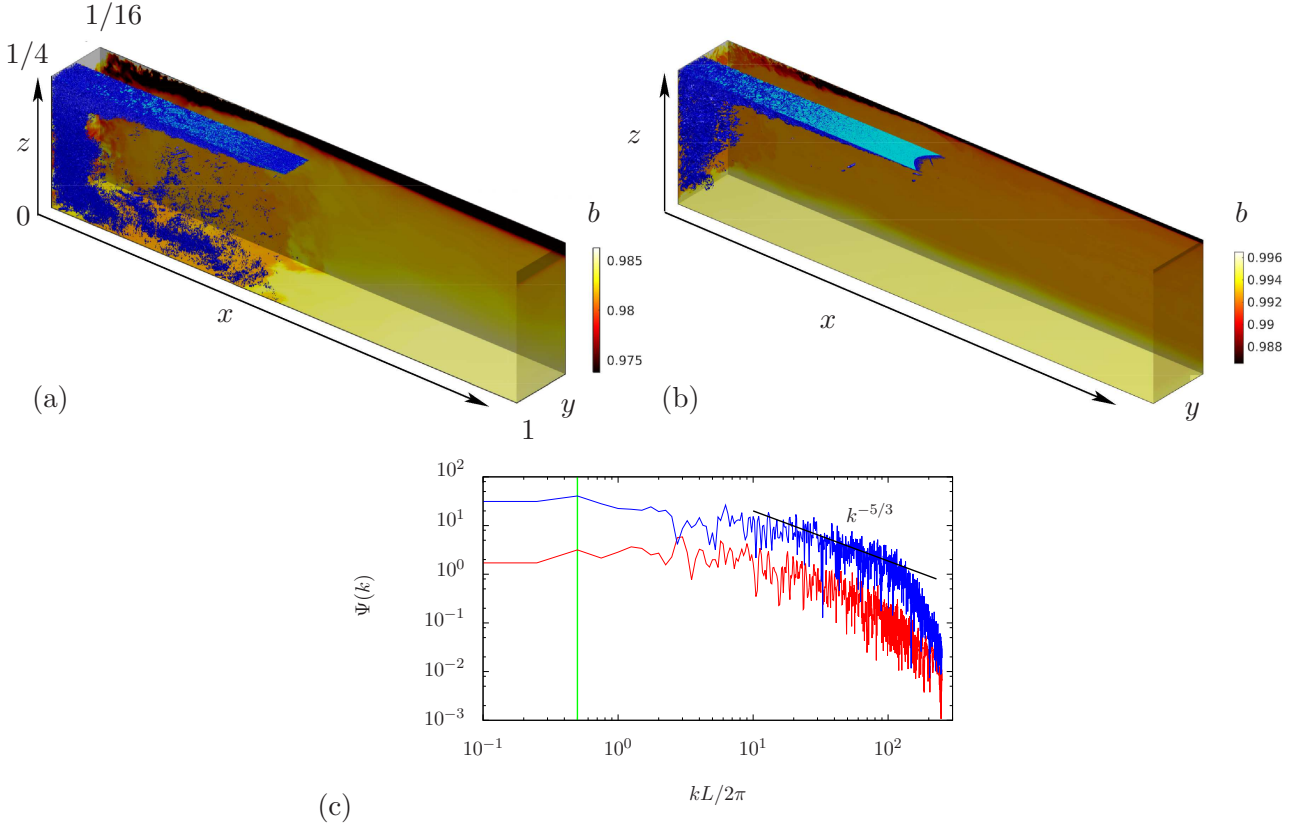}%
\put(-490,130){$1/4$}\put(-460,145){$1/16$}\put(-486,69){$0$}\put(-295,-12){$1$}
\put(-486,97){$z$}\put(-410,32){$x$}\put(-280,-2){$y$}
\put(-255,100){$z$}\put(-182,30){$x$}\put(-44,-3){$y$}
\put(-480,0){(a)}\put(-242,0){(b)}
\put(-270,70){$b$}\put(-25,70){$b$}\\
\centering
(c)\hspace{-2mm}\scalebox{0.7}{\input{spectra_2}}%
\vspace{-3mm}
\caption{Snapshot of the iso-contours of $\Lambda_2=-\rm{Pr}^{-1}/8$ criteria (blue) and buoyancy $b$ (background) at (a) $Ra=6.4\times 10^{13}$, $Pr=0.01$ showing the new regime ($II_l$) and (b) $\rm{Ra}=6.4\times 10^{13}$, $Pr=1$ corresponding to the Hughes' regime \cite{Hughes07} ($II_u$).(c) Turbulent spectra $\Psi(k)$ at $Ra=6.4\times 10^{13}$ at $\rm{Pr}=0.05$ (red) and $\Psi(k)/2$ at $\rm{Pr}=1$ (blue) for the same $\rm{Ra}$. The black line shows the $k^{-5/3}$ turbulence cascade whereas the green vertical line shows the wavenumber of the forcing.}
\label{3D}
\end{figure*}
%
%
The Prandtl number is given by $\rm{Pr}=\nu/\kappa$ where $\nu$ and $\kappa$ are the viscous and stratifying agent's diffusion coefficients.
The Navier-Stokes equations are solved on a Cartesian grid, stretched near the upper boundary using second-order finite volumes while the pressure is solved using a standard projection method. Laminar and turbulent flow solutions were integrated in time using an explicit second order Adams-Bashforth type-scheme.
The Rayleigh number is defined such that ${\rm Ra}=\Delta L^3/(\nu\kappa)$ where $L$ is the horizontal length scale of the domain and $\Delta=-g(\rho_{max}-\rho_{min})/\rho_{min}$. The computational domain is a parallelepiped of aspect ratio $\Gamma=4$ with dimensions $[L,W,H]=[1,1/8,1/4]$ where $W$ is the width of the computational domain \cite{Scotti08}. A buoyancy profile is imposed at the surface $z=H$ where $H$ is the height of the domain using a buoyancy profile such that $b(x)|_{z=H}=(1+\tanh(9.5x))/2$ (see ref.\cite{PassaggiaSW17}). 
Since we are interested in turbulence dominated regimes where the scaling is not determines by the buoyancy forcing profile, nor the aspect ratio \cite{ShishkinaGL16,Sheard2011}, nor the ype of boundary condition, free-slip boundary conditions are used for the velocity on the upper, lower and end walls at $x=\pm L/2$ (see ref.\cite{ScottiW11}) while the domain is assumed periodic in the transverse direction $y$. This is in contrast with Shishkina \& Wagner\cite{ShishkinaW16} where they used no-slip boundary conditions and end walls in the transverse direction. 
Our approach avoids the numerical difficulties involved with resolving the no-slip BL and a finite domain in the transverse direction. Instead we privilege numerical efficiency and report results for $\rm{Ra}$ up to $6.4\times10^{13}$ for a wide range of Prandtl numbers.

\begin{figure}[t!]
\centering
\hspace{-2mm}\scalebox{1.25}{\scriptsize\input{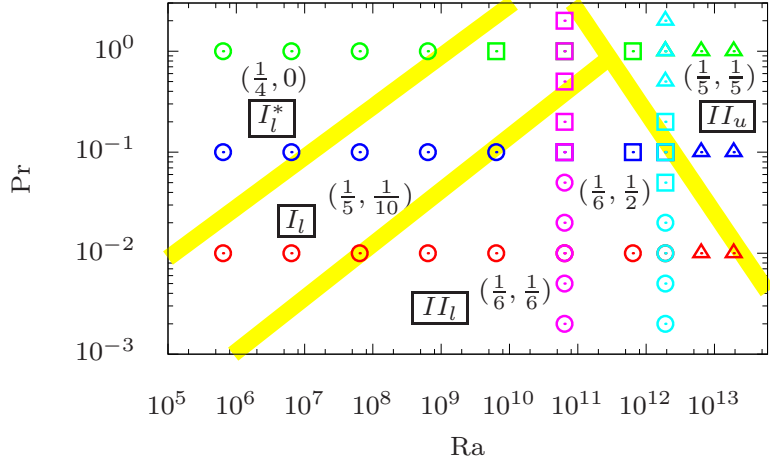}}%
\vspace{-10mm}
\caption{(a) Sketch of the phase diagram in the $(\rm{Ra}, \rm{Pr})$ plane for the
laminar regimes $I_l$ and  $I^*_l$ together with the turbulent scalings $II_{l}$ with the conducted DNS.
The yellow stripes shows the transition from $I^*_l$ to $I_{l}$, and $I_l$ to $II_{l}$, with a
slope $Pr\approx Ra^{1/2}$. The transition from $II_l$ to $II_u$ with a
slope $Pr\approx Ra^{-1}$.  Symbols reflect the computational meshes in $(x,y,z)$, used in the DNS: $512\times256\times256$ (circle), $1024\times384\times128$ (squares), and $2048\times256\times256$ (squares). The values ($\alpha,\beta$) in each region provide the exponents $\rm{Nu}\sim \rm{Ra}^{\alpha}\rm{Pr}^{\beta}$ measured in the DNS and derived in the theory. 
}
\label{Ra_Pr}
\end{figure}

The turbulent scalings for momentum and buoyancy transport are computed using Direct Numerical Simulations (DNS) in the range $\rm Ra=[6.4\times10^5,6.4\times10^{13}]$ and $0.002\leq \rm Pr\leq 2$. 
For $\rm Ra<10^8$ and $0.5\leq \rm Pr\leq 2$, the HC flows are steady \cite{ShishkinaW16,PassaggiaSW17}.
With increasing $\rm{Ra}$ and/or decreasing values of $\rm{Pr}$, HC flows become increasingly unsteady, leading to turbulence (as shown in figure \ref{3D}(c)) and the mesh size is increased in order to resolve the Kolmogorov length scale (see ref.\cite{ScottiW11} for details about turbulent HC). Mesh sizes are reported in Fig.\ref{Ra_Pr}(b) in the $(\rm{Ra},\rm{Pr})$ plane along with the different regimes reported later in this manuscript. Note that turbulence in HC for moderate values of $\rm{Pr}$ is confined to a narrow region located under the cooling/heavy boundary consisting of the plume and the BL where the fluid is statically unstable (cf. Fig.\ref{3D}) \cite{Gayen14,ScottiW11}. Decreasing values of $\rm{Pr}$ increases the volume of fluid subject to turbulence (see Fig.\ref{3D}) and extends the depth of the circulation.\\

\begin{figure*}[t!]
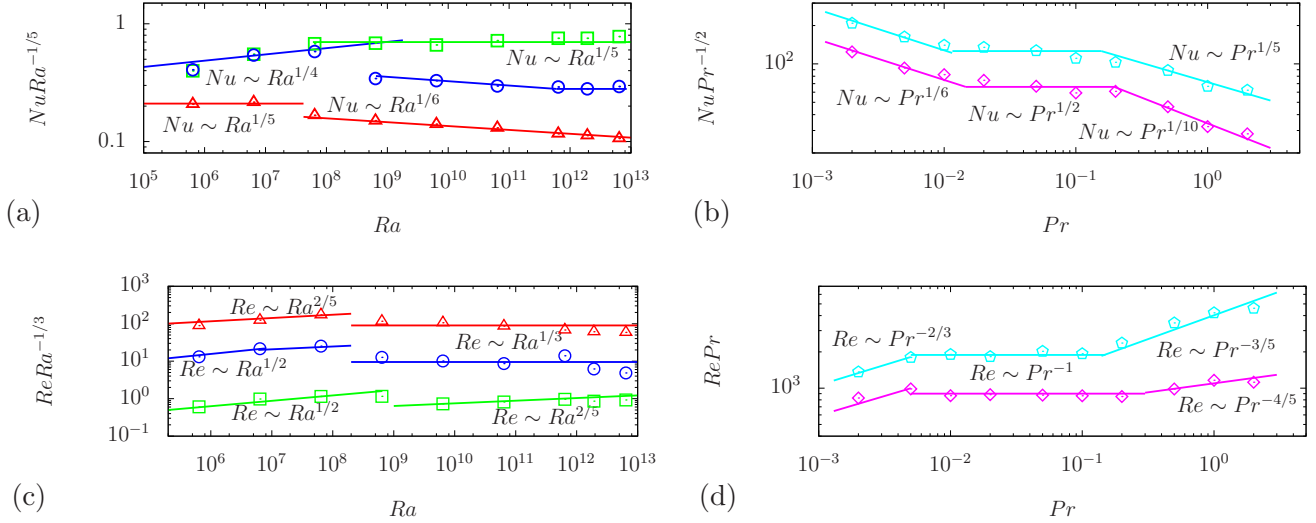

\centering
\begin{minipage}[b]{\linewidth}
\scalebox{0.7}{\large\input{NuRa1e02_Ra_2}}
\scalebox{0.7}{\large\input{NuPr-05_Pr_2}}
\put(-510,10){(a)}\put(-250,10){(b)}\\
\vspace{-6mm}
\scalebox{0.7}{\large\input{ReRa-03_Ra_2}}
\scalebox{0.7}{\large\input{RePr06_Pr_2}}
\put(-510,10){(c)}\put(-250,10){(d)}
  \end{minipage}
\vspace{-3mm}
\caption{(a),(c) $\rm{Ra}$ dependencies and (b),(d) $\rm{Pr}$ dependencies of (a),(b) the Nusselt number and (c),(d) the Reynolds number, as obtained in the DNS for (a),(c) $\rm{Pr}=1$ (squares), $\rm{Pr}=0.1$ (circles), $\rm{Pr}=0.01$ (triangles) and for (b),(d) $\rm{Ra}=6.4\times10^{10}$  (diamonds) and $\rm{Ra}=1.92\times10^{12}$ (pentagones). $\rm{Pr}$ dependence of $\bar{\epsilon_u}$ with $\rm{Re}$ (d).
The DNS results support the scaling in the regime $I_l$ (solid lines) [Eqs. (\ref{lam_scal}a) and (\ref{lam_scal}b)], transition to $II_l$ (dotted lines) [Eqs. (\ref{NuRe1/6}a),(\ref{NuRe1/6}b) and \ref{NuRe1/6corr}),(\ref{NuRe1/6corr})], transition to $II_u$ (dotted lines) [Eqs. (\ref{NuRe1/5h}a) and (\ref{NuRe1/5h}b)].}
\label{Nu_Re}
\end{figure*}

\begin{figure}[t!]
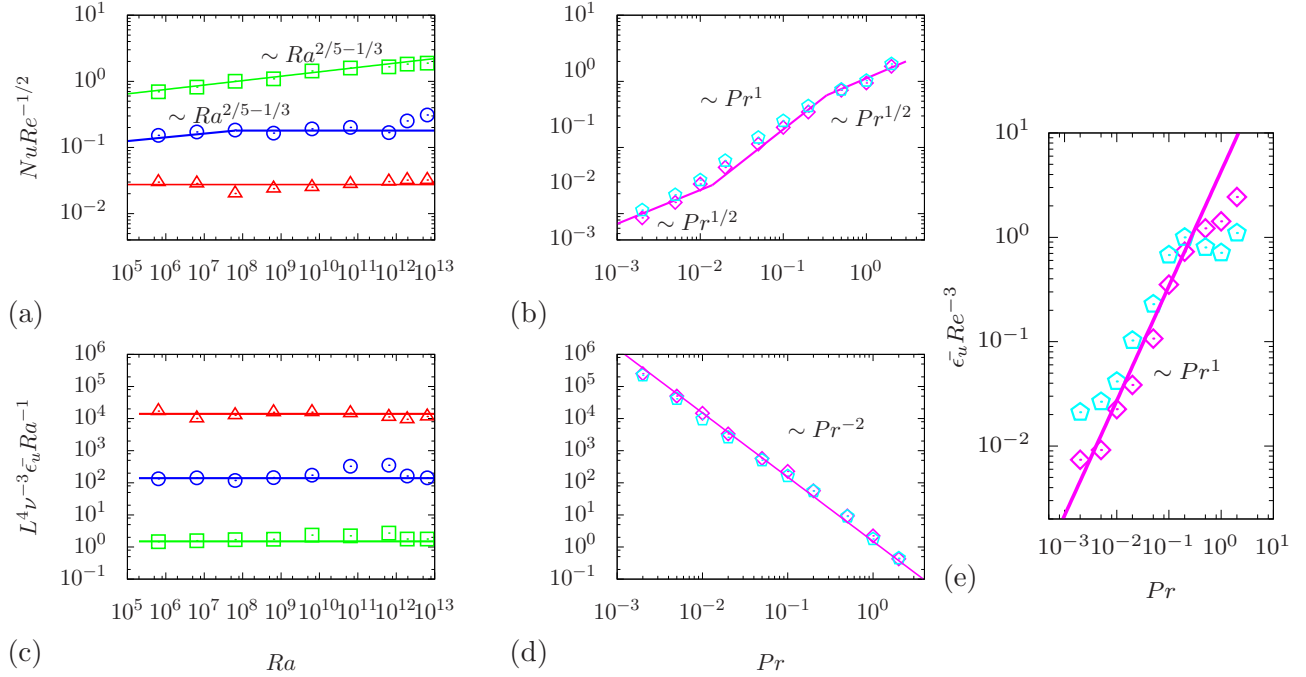


\begin{minipage}[b]{0.77\linewidth}
\hspace{-6mm}\scalebox{0.8}{\input{NuRe05_Ra_2}}%
\hspace{-6mm}\scalebox{0.8}{\input{NuRe05_Pr_2}}%
\put(-370,10){(a)}\put(-180,10){(b)}
\vspace{-5mm}

\hspace{-6mm}\scalebox{0.8}{\input{L4nu-3epsilonRa-1_Ra_2}}%
\hspace{-6mm}\scalebox{0.8}{\input{L4nu-3epsilonRa-1_Pr_2}}%
\put(-370,10){(c)}\put(-180,10){(d)}
\end{minipage}
\hspace{-14mm}
\begin{minipage}[b]{0.22\linewidth}
\hspace{-0mm}\scalebox{1.1}{\scriptsize\input{epsilonRe-3_Pr_2}}
\put(-145,10){(e)}
\vspace{10mm}
  \end{minipage}
\vspace{-3mm}
\caption{(a),(c) $\rm{Ra}$ dependences and (b),(d) $\rm{Pr}$ dependences of
(a),(b) $\rm{Nu}\rm{Re}^{-1/2}$ and (c),(d) $L^4\nu^{-3}\bar{\epsilon_u}\rm{Ra}^{-1}$, 
as obtained in the DNS for (a),(c) $\rm{Pr}=1$ (squares), $\rm{Pr}=0.1$ (circles), 
$\rm{Pr}=0.01$ (triangles) and for (b),(d) $Ra=10^9$ 
(diamonds) and $\rm{Ra}=2\times10^{10}$ (pentagones). The upper
figures support (\ref{Nu}) and (\ref{NuRe1/6corr}b), while the lower figures illustrate (\ref{epsu1}).
The correction in (e) supports the scaling (\ref{epsub2}).}
\label{Re_eps}
\end{figure}

The dependences of $\rm{Nu}$ and $\rm{Re}$ with respect to $\rm{Ra}$ and $\rm{Pr}$ are summarized in Fig. \ref{Nu_Re}(a-d). For all values of $\rm{Pr}$ and $\rm{Ra}$, the Nusselt number  $\rm{Nu}\sim Ra^{\alpha}$ [see Fig. \ref{Nu_Re}(a)] transitions from:
\begin{itemize}
\item the enhanced laminar scaling $\alpha=1/4$ for low $\rm{Ra}$,
\item the classical laminar scaling $\alpha=1/5$ for higher $\rm{Ra}$, 
\item A new $\alpha=1/6$ exponent for small $\rm{Pr}$, and
\item the entrainment-type regime $\alpha=1/5$ at even higher $\rm{Ra}$.
\end{itemize}
 The magnitude of the large-scale flow is given by $\rm{Re}={(\overbar{\mathbf{u}\cdot\mathbf{u}})}^{1/2}L/\nu$ where the overbar denotes the spatio-temporal average over the computational domain. We observe the laminar scalings 
$Re\sim \rm{Ra}^{\gamma}$ with $\gamma=1/2$ (see ref. \cite{ShishkinaW16}) and $\gamma=2/5$ (see ref.\cite{Rossby65}). At higher $\rm{Ra}$, the new scaling $\gamma=1/3$ is also observed and changes back to $\gamma=2/5$ (see ref.\cite{Hughes07}) [Fig. \ref{Nu_Re}(c)].
The low $\rm{Pr}$ values show a dependence on $\rm{Nu}\sim Pr^{\beta}$ with:
\begin{itemize}
\item $\beta=1/2$ for higher $\rm{Pr}$ and low $\rm{Ra}$ (see ref. \cite{ShishkinaW16}),
\item $\beta=1/10$ for $\rm{Ra}<10^{11}$ see (see ref.\cite{Rossby65}),  
\item the new exponent $\beta=1/6$ at low $\rm{Pr}$, and
\item $\beta=1/5$ for $\rm{Ra}>5\times10^{11}$ see (see ref.\cite{Hughes07}). 
\end{itemize}
The Reynolds number dependence $\rm{Re} \sim \rm{Pr}^{\delta}$ with $\delta=-2/3$ for the smaller value of $\rm{Pr}$, then $\delta=-1$ for $10^{-2}\lesssim\rm{Pr}\lesssim0.2$
changes to $\delta=-4/5$ for increasing $\rm{Ra}$ at all $\rm{Pr}$ and increases at high $\rm{Ra}$ to the HGM scaling $\delta=-3/5$ for the larger values of $\rm{Ra}$ [Fig.\ref{Nu_Re}(d)].\\


The scalings obtained from the DNS can be derived from eq. (\ref{NS}c) for the steady thermal BL equation which writes \cite{GL00,ShishkinaWH14}
\begin{equation}
u b_x + v b_z = \kappa b_{zz},
\end{equation}
and reduces to $U\Delta/L = \kappa \Delta/\lambda_b^2$ where $\lambda_b$ is the thickness of the thermal BL, which scales as $\lambda_b \sim \rm{Nu}^{-1}$. Combining the above reduces to
\begin{equation}
\rm Nu=Re^{1/2}Pr^{1/2},
\label{Nu}
\end{equation}
and provides a relation tying $\rm{Nu}$, $\rm{Re}$ and $\rm{Pr}$. This result is supported by our DNS of laminar to turbulent HC [Fig. \ref{Re_eps}(a)], for small $\rm Pr$. A small correction was found for $\rm{Pr}=1$ and can also be observed in the laminar DNS described in ref.\cite{ShishkinaW16}. Also note that the Prandtl number dependence on (\ref{Nu}) is modified from $\sim\rm Pr^{1/2}$ to $\sim\rm Pr^{1}$ for $10^{-2}\lesssim\rm{Pr}\lesssim 0.2$ and is associated with the modification of the flow [see Fig. \ref{3D}(a,b)] which will be explained later.\\

Paparella and Young (PY) \cite{PaparellaY02} showed that the amount of kinetic energy dissipation in HC is bounded.
The time and volume average of eq. (\ref{NS}c) in combination with eq. (\ref{NS}b) give $\langle \partial b/\partial z\rangle_{z=H}=0$ where $\langle\rangle_{z=H}$ denotes the surface and time average at $z=H$. Combining the time average of eq. (\ref{NS}c) with the PY constraint
(i.e. $\langle wb\rangle_z=\kappa\langle \partial b/\partial z\rangle_z$) (see ref.\cite{PaparellaY02}), and integrating over $z$ leads to
\begin{equation}
\overbar{wb}\leq\kappa(\langle b\rangle_{z=H}-\langle b\rangle_{z=0})/H = B(\Gamma/2)\kappa\Delta/L,
\end{equation}
where $1<B<0$ is an arbitrary constant \cite{ShishkinaW16}. Taking the kinetic energy equation, time and volume averaging,
\begin{equation}
\overbar{\epsilon_u} \;=\; \overbar{wb} \;\leq\; B(\Gamma/2)\nu^{3}L^{-4}\rm{Ra}\rm{Pr}^{-2},
\label{epsu2}
\end{equation}
which is supported for all regimes. SGL recast this argument in a spatio-temporal volume-averaged kinetic energy dissipation rate $\overbar{\epsilon_u}\equiv\nu\sum_{i,j}(\partial u_j/\partial x_i)^2$ which in the case of a BL-dominated regime, matches the dissipation in the boundary layer $\overbar{\epsilon_u}\sim(\nu U^2)/ (\lambda_u L)$ where $\lambda_u$ is the thickness of the viscous BL. 
Together with the scaling for the BL thickness such that $\lambda_u \sim Re^{-1/2}$ , the scaling for the mean dissipation in the particular case of laminar BL\cite{Landau87} is
\begin{equation}
\overbar{\epsilon_u}\sim\nu^3L^{-4}{\rm Re}^{5/2}.
\label{epsu1}
\end{equation}
%
Combining (\ref{Nu}), (\ref{epsu2}) and (\ref{epsu1}), one recovers the laminar scaling \cite{Rossby65,Gayen14,ShishkinaGL16}
\begin{subeqnarray}
\rm{Re} &\sim& \rm{Ra}^{2/5}\rm{Pr}^{-4/5}, \\
\rm{Nu} &\sim& \rm{Ra}^{1/5}\rm{Pr}^{1/10}.
\label{lam_scal}
\end{subeqnarray}
By analogy to the notation in the GL theory for RBC \cite{GL00,ShishkinaGL16}, this scaling regime is denoted as $I_l$, where the subscript $l$ stands for low-$\rm{Pr}$ fluids. 
With decreasing $\rm{Pr}$ and/or increasing $\rm{Ra}$, the bulk dynamics is driven by the large-scale overturning flow whose horizontal length scale is $L$. In this case, it is the large-scale velocity $U$ which drives the dissipation of kinetic energy and the latter is given by
\begin{equation}
\overbar{\epsilon_u} \sim \nu^{3}L^{-4}\rm{Re}^3.
\label{epsub}
\end{equation}

From (\ref{Nu}), (\ref{epsu2}) and (\ref{epsub}), it follows that low-$\rm{Pr}$ HC exhibits dependences of the form
\begin{subeqnarray}
\rm{Re}&\sim&\rm{Ra}^{1/3}\rm{Pr}^{-2/3}, \\
\rm{Nu}&\sim&\rm{Ra}^{1/6}\rm{Pr}^{1/6},
\label{NuRe1/6}
\end{subeqnarray}
where this scaling regime is denoted as $II_l$ [see Fig. \ref{Ra_Pr}(b) and ref.\cite{ShishkinaGL16}].
Note that these scalings are only observed for $\rm{Pr}\lesssim10^{-2}$ where the large-scale flow spans the full depth of the domain [see Fig. \ref{3D}(a)]. For intermediate $\rm{Pr}$, the size of the turbulent overturning flow decreases with increasing $\rm{Pr}$, from a deep to a shallow region [see Fig. \ref{3D}(a-b)]. Similarly, the dissipation increases linearly with increasing $\rm{Pr}$ and eq. (\ref{Nu}) together with eq. (\ref{epsub}) can be modified to account for that transitional regime such that
\begin{subeqnarray}
\overbar{\epsilon_u} &\sim& \nu^{3}L^{-4}\rm{Re}^3\rm{Pr}, \\
\rm{Nu} &\sim& \rm{Re}^{1/2}\rm{Pr}
\label{epsub2}
\end{subeqnarray}
which is verified empirically in our DNS [see Fig. \ref{Nu_Re}(e) and Fig. \ref{Re_eps}(b)]. Combining eq. (\ref{epsu2}), and eq. (\ref{epsub2}a-b)) provides a correction for this $\rm{Pr}$ transition in the $II_l$ regime
\begin{subeqnarray}
\rm{Re}&\sim&\rm{Ra}^{1/3}\rm{Pr}^{-1}, \\
\rm{Nu}&\sim&\rm{Ra}^{1/6}\rm{Pr}^{1/2},
\label{NuRe1/6corr}
\end{subeqnarray}
found in the range $10^{-2}\lesssim \rm{Pr}\lesssim 0.2$ [see Fig. \ref{Nu_Re}(b,d)].
Further increasing $\rm{Pr}$ and for low $\rm{Ra}$, the BL thickness $\lambda_u$ saturates
and we observe the laminar scaling  reported by Shishkina \& Wagner \cite{ShishkinaW16}
\begin{subeqnarray}
\rm{Re}&\sim&\rm{Ra}^{1/2}\rm{Pr}^{-1}, \\
\rm{Nu}&\sim&\rm{Ra}^{1/4}\rm{Pr}^{0},
\label{NuRe1/4}
\end{subeqnarray}
denoted as $I^*_l$ [see Fig. \ref{Ra_Pr}(b)].
Increasing $\rm{Ra}$ and at high $\rm{Pr}$, the dynamics are driven by the plume, detached from the bottom [see Fig. \ref{3D}(b)]. This particular case was theorized by Hughes {\it et al.}\cite{Hughes07} with a plume model inside a filling box. Here we recast their model according to the SGL theory (i.e. see the plume model definition eq. (2.15)-(2.20) in  ref.\cite{Hughes07}) and the dissipation in the boundary layer writes
\begin{equation}
\overbar{\epsilon_{u,Pl}} \sim \nu^{3}L^{-4}\rm{Re}^{5/2}\rm{Pr}^{-1/2},
\label{epsuh}
\end{equation}
where the dissipation now scales with the thickness of the thermal layer, not the kinetic BL and is given by $\overbar{\epsilon_{u,Pl}}\sim\nu U^2/(\lambda_b L)$. Combining (\ref{Nu}), (\ref{epsu2}) and (\ref{epsuh})
\begin{subeqnarray}
\rm{Re}&\sim&\rm{Ra}^{2/5}\rm{Pr}^{-3/5}, \\
\rm{Nu}&\sim&\rm{Ra}^{1/5}\rm{Pr}^{1/5},
\label{NuRe1/5h}
\end{subeqnarray}
which is denoted as $II_u$ [see Fig. \ref{Ra_Pr}(b)].\\

The slope of the transition regions in the $(\rm{Ra},\rm{Pr})$ plane, between the laminar regimes $I_l$, $II_l$, $II_u$ and $I_l^{\infty}$, is determined 
by matching the Reynolds numbers in these neighboring regimes \cite{GL00,ShishkinaGL16}. Thus, from eqs. (\ref{lam_scal}) and (\ref{NuRe1/4}), we obtain the slope of the transition region between the regimes $I_l$ and $II_l$, which is $\rm{Pr}\sim\rm{Ra}^{1/2}$  and $\rm{Pr}\sim\rm{Ra}^{-1}$ between $II_l$ and $II_u$ \cite{ShishkinaW16}.  Each transition region is highlighted by a yellow line in the $(\rm{Ra},\rm{Pr})$ plane [see Fig. \ref{Ra_Pr}(b)] and is
estimated from the DNS data, by considering the changes in the $\rm{Nu}(\rm{Ra},\rm{Pr})$ and $\rm{Re}(\rm{Ra},\rm{Pr})$ dependencies. 
Note that the transition is not smooth and suggests that a bifurcation takes place when transitioning from the $I_l$ to $II_u$ regime, and can be affected by the geometry of particular HC setups (as previously suggested in ref.\cite{ShishkinaW16}).\\

In conclusion, we report evidences of a new regime in turbulent simulation of horizontal convection based on scaling arguments at low $\rm{Pr}$. The transition occurs from $\rm{Re} \sim \rm{Ra}^{2/5}\rm{Pr}^{-4/5}, \rm{Nu} \sim \rm{Ra}^{1/5}\rm{Pr}^{1/10}$ to 
$\rm{Nu}\sim \rm{Ra}^{1/6}\rm{Pr}^{1/6}, \rm{Re}\sim \rm{Ra}^{1/3}\rm{Pr}^{-2/3}$ at low $\rm{Pr}$ and sufficiently large $\rm{Ra}$ and then $\rm{Re} \sim \rm{Ra}^{2/5}\rm{Pr}^{-3/5}, \rm{Nu} \sim \rm{Ra}^{1/5}\rm{Pr}^{1/5}$.
Our results, integrate previous evidence from Shishkina \& Wagner \cite{ShishkinaW16} and the model of Hughes {\it et al.} (see ref. \cite{Hughes07}) in the SGL theory of HC (see ref. \cite{GL00,ShishkinaGL16}). 
The transition to the turbulent limiting regime denoted as $IV_u$ or the ultimate regime $IV_l$ have yet to be observed (see ref. \cite{ShishkinaGL16}). It is therefore of particular interest to attain Rayleigh number of $\rm{Ra}\approx10^{15}$ to determine whether core driven turbulent HC regimes can be reached, as suggested in ref.\cite{ShishkinaGL16}.\\

The authors acknowledge the support of the National Science Foundation Grant Number OCE--1155558 and OCE--1736989.

\bibliographystyle{plain}

\bibliography{bib}

\providecommand{\noopsort}[1]{}\providecommand{\singleletter}[1]{#1}%
\begin{thebibliography}{10}

\bibitem{alboussiere:12}
T.~Alboussiere, R.~Deguen, and M.~Melzani.
\newblock Melting-induced stratification above the earth's inner core due to
  convective translation.
\newblock {\em Nature}, 466:744--747, 2010.

\bibitem{bloxham:91}
J.~Bloxham and A.~Jackson.
\newblock Fluid flow near the surface of earth's outer core.
\newblock {\em Rev. Geophys.}, 29(1):97--120, 1991.

\bibitem{Gayen14}
B.~Gayen, R.~W. Griffiths, and G.~O. Hughes.
\newblock Stability transitions and turbulence in horizontal convection.
\newblock {\em J. Fluid Mech.}, 751:698--724, 7 2014.

\bibitem{GL00}
S.~Grossmann and D.~Lohse.
\newblock Scaling in thermal convection: a unifying theory.
\newblock {\em J. Fluid Mech.}, 407:27--56, 2000.

\bibitem{hughes2008horizontal}
G.~O. Hughes and R.~W. Griffiths.
\newblock Horizontal convection.
\newblock {\em Annu. Rev. Fluid Mech.}, 40:185--208, 2008.

\bibitem{Hughes07}
G.~O. Hughes, R.~W. Griffiths, J.~C. Mullarney, and W.~H. Peterson.
\newblock A theoretical model for horizontal convection at high rayleigh
  number.
\newblock {\em J. Fluid Mech.}, 581:251--276, 2007.

\bibitem{Knopoff63}
L.~Knopoff.
\newblock Horizontal convection in the earth's mantle: A mechanism for
  strike-slip faulting.
\newblock {\em Science}, 140(3565):383--383, 1963.

\bibitem{Landau87}
L.~D. Landau and E.~M. Lifschitz.
\newblock {\em Statistische Physik}.
\newblock Akademie-Verlag, 1987.

\bibitem{PaparellaY02}
F.~Paparella and W.~R. Young.
\newblock {Horizontal convection is non-turbulent}.
\newblock {\em J. Fluid Mech.}, 466:205--214, 2002.

\bibitem{PassaggiaSW17}
P.-Y. Passaggia, A.~Scotti, and B.~L. White.
\newblock Transition and turbulence in horizontal convection: linear stability
  analysis.
\newblock {\em J. Fluid Mech.}, 821:31--58, 2017.

\bibitem{Roche02}
P.-E. Roche, B.~Castaing, B.~Chabaud, and B.~H{\'e}bral.
\newblock Prandtl and rayleigh numbers dependences in rayleigh-b{\'e}nard
  convection.
\newblock {\em Europhys. Lett.)}, 58(5):693, 2002.

\bibitem{Rossby65}
H.~T. Rossby.
\newblock On thermal convection driven by non-uniform heating from below: an
  experimental study.
\newblock {\em Deep-Sea Res.}, 12:9--16, 2 1965.

\bibitem{Scotti08}
A.~Scotti.
\newblock A numerical study of the frontal region of gravity currents
  propagating on a free-slip boundary.
\newblock {\em Theo. Comput. Fluid Dyn.}, 22(5):383, 2008.

\bibitem{ScottiW11}
A.~Scotti and B.~L. White.
\newblock {Is Horizontal convection really "non turbulent"?}
\newblock {\em Geophys. Res. Lett.}, 38:L21609, 2011.

\bibitem{Sheard2011}
G.~J. Sheard and M.~P. King.
\newblock Horizontal convection: effect of aspect ratio on rayleigh number
  scaling and stability.
\newblock {\em App. Math Model.}, 35(4):1647--1655, 2011.

\bibitem{ShishkinaGL16}
O.~Shishkina, S.~Grossman, and D.~Lohse.
\newblock Heat and momentum transport scalings in horizontal convection.
\newblock {\em Geophys. Res. Lett.}, 43(3):1219--1225, 2016.

\bibitem{ShishkinaW16}
O.~Shishkina and S.~Wagner.
\newblock Prandtl-number dependence of heat transport in laminar horizontal
  convection.
\newblock {\em Phys. Rev. Lett.}, 116(2):024302, 2016.

\bibitem{ShishkinaWH14}
O.~Shishkina, S.~Wagner, and S.~Horn.
\newblock Influence of the angle between the wind and the isothermal surfaces
  on the boundary layer structures in turbulent thermal convection.
\newblock {\em Phys. Rev. E}, 89(3):033014, 2014.

\bibitem{shishkina2017scaling}
Olga Shishkina, Mohammad~S Emran, Siegfried Grossmann, and Detlef Lohse.
\newblock Scaling relations in large-prandtl-number natural thermal convection.
\newblock {\em Physical review fluids}, 2(10):103502, 2017.

\bibitem{takehiro:11}
S.-I. Takehiro.
\newblock Fluid motions induced by horizontally heterogeneous joule heating in
  the earth's inner core.
\newblock {\em Phys. Earth Planet. Inter.}, 184(3):134--142, 2011.

\bibitem{Taylor13}
M.~F. Taylor, K.~E. Bauer, and D.~M. McEligot.
\newblock Internal forced convection to low-prandtl-number gas mixtures.
\newblock {\em Int. J. Heat Mass Trans.}, 31(1):13 -- 25, 1988.

\end{thebibliography}

\end{document}